\documentclass[aps,prb,twocolumn,superscriptaddress,showpacs]{revtex4}

\usepackage{graphicx}
\usepackage{amsmath}
\usepackage[applemac]{inputenc}
\usepackage{color}

\newcommand{\DyTi}{Dy$_2$Ti$_2$O$_7$}

\usepackage[T1]{fontenc}
\usepackage{longtable}

\begin{document}

\title{Dynamic behavior of magnetic avalanches in the spin-ice compound \DyTi.}

\author{M. J. Jackson}\altaffiliation[]{current address: Faculty of Mathematics and Physics, Charles University in Prague, Ke Karlovu 3, 121 16, Praha 2, Czech Republic}
\affiliation{Institut N\'eel, CNRS \& Universit\'e Joseph Fourier, BP 166, 38042 Grenoble Cedex 9, France.}
\author{E. Lhotel}
\email[]{elsa.lhotel@neel.cnrs.fr}
\affiliation{Institut N\'eel, CNRS \& Universit\'e Joseph Fourier, BP 166, 38042 Grenoble Cedex 9, France.}
\author{S. R. Giblin} \affiliation{School of Physics and Astronomy, Cardiff University, Cardiff, CF24 3AA, United Kingdom.}
\author{S. T. Bramwell} \affiliation{London Centre for Nanotechnology and Department of Physics and Astronomy, University College London, London WC1H 0AH, United Kingdom.}
\author{D. Prabhakaran} \affiliation{Department of Physics, Clarendon Laboratory, University of Oxford, Oxford OX13PU, United Kingdom.}
\author{K. Matsuhira}\affiliation{Kyushu Institute of Technology, Kitakyushu 804-8550, Japan.}
\author{Z. Hiroi} \affiliation{Institute for Solid State Physics, University of Tokyo, Kashiwa, Chiba 277-8581, Japan.}
\author{Q. Yu} \affiliation{Institut N\'eel, CNRS \& Universit\'e Joseph Fourier, BP 166, 38042 Grenoble Cedex 9, France.}
\author{C. Paulsen} \affiliation{Institut N\'eel, CNRS \& Universit\'e Joseph Fourier, BP 166, 38042 Grenoble Cedex 9, France.}


\begin{abstract}
Avalanches of the magnetization, that is to say an abrupt reversal of the magnetization at a given field, have been previously reported in the spin-ice compound Dy$_{2}$Ti$_{2}$O$_{7}$. This out-of-equilibrium process, induced by magneto-thermal heating, is quite usual in low temperature magnetization studies. A key point is to determine the physical origin of the avalanche process.
In particular, in spin-ice compounds, the origin of the avalanches might be related to the monopole physics inherent to the system. We have performed a detailed study of the avalanche phenomena in three single crystals, with the field oriented along the [111] direction, perpendicular to [111] and along the [100] directions. We have measured the changing magnetization during the avalanches and conclude that avalanches in spin ice are quite slow compared to the avalanches reported in other systems such as molecular magnets. Our measurements show that the avalanches trigger after a delay of about 500 ms and that the reversal of the magnetization then occurs in a few hundreds of milliseconds. These features suggest an unusual propagation of the reversal, which might be due to the monopole motion. The avalanche fields seem to be reproducible in a given direction for different samples, but they strongly depend on the initial state of magnetization and on how the initial state was achieved. 
\end{abstract}

\maketitle

\section{Introduction}
Geometrically frustrated magnetic systems provide a large variety of unusual magnetic ground states \cite{Lacroix}. Among these, the pyrochlore compounds (formula A$_2$B$_2$O$_7$, where A is a magnetic rare-earth, and B a transition metal), in which the magnetic ions form a network of corner-sharing tetrahedra, is a very rich family \cite{Gardner10}. In particular, when A=Dy or Ho and B=Ti, the magnetic moments have a strong Ising-like anisotropy and are directed along the 
$\langle111\rangle$ local axis of the tetrahedra. The effective ferromagnetic interaction (resulting from the dipolar interaction and a weak antiferromagnetic exchange \cite{Siddharthan99, denHertog00}) induces a disordered low temperature state called the spin ice state\cite{Harris97}. In this state, the local spin arrangement on a tetrahedron obeys the so-called ``ice rules'', in which two spins point into the tetrahedron and two spins point out of the tetrahedron (``two in - two out'' state). Magnetization measurements have shown that below 0.7 K, these states become frozen on experimental timescales \cite{Snyder04}. Recently, it was shown that excitations from the ground state (``three in - one out'' or ``three out - one in'') carry an effective magnetic charge and can be described as de-confined magnetic monopoles coupled by a Coulomb interaction\cite{Castelnovo08, Ryzhkin05}.

In the presence of an applied magnetic field, the degeneracy of the spin-ice ground state is partially or totally lifted. Depending on the direction of the magnetic field with respect to the spin ice crystal, very different field-induced states are predicted \cite{Harris98}. The underlying physics of these field induced states has been shown to be very rich. 
Along the [111] direction of the crystal, the pyrochlore can be viewed as the stacking of alternate triangular and kagom\'e planes. When the field is applied along this direction, the magnetization curve shows a plateau at about 2000 Oe which has been attributed to the so-called kagom\'e ice state \cite{Matsuhira02, Sakakibara03, Tabata06}: in this state, the spins of the triangular planes are aligned with the magnetic field and the kagom\'e spins respect the ``two in - one out'' or ``one in - two out'' rule in their triangle. When the field is further increased, a liquid-gas type transition \cite{Sakakibara03, Fennell07} has been observed. Kasteleyn transitions have been predicted and observed when the field is slightly deviated from the [111] direction \cite{Moessner03, Fennell07} or is along the [100] direction \cite{Jaubert08}. When the field is along the [110] direction, the system can be considered as two independent sets of spin-chains, paramagnetic and ferromagnetic \cite{Fennell05}. The magnetic and specific heat behavior in more other configurations of the field have evidenced the existence of further neighbor interactions \cite{Ruff05, Higashinaka05}. 

Magnetic avalanches in the spin ice compound \DyTi~were first detected by neutron scattering~\cite{Fennell05}. More recently they have been studied by Slobinsky {\it et al.} using bulk magnetometry with a field applied along the [111] direction~\cite{Slobinsky10}. Their signature has also been observed by ultrasound measurements \cite{Erfanifam11}. 
Such magnetic avalanches, that is to say fast reversals of the magnetization at a given magnetic field, are often present in magnets at low temperature. They have been reported in various materials, such as manganites \cite{Macia07}, rare-earth metallic glasses \cite{Hadjipanayis81}, spin-glasses \cite{Prejean80}, heavy fermions \cite{Lhotel04, Marcano07} or molecular magnets \cite{Paulsen95, Lhotel08} for example. In some cases, in particular in the presence of magnetic disorder, they can be considered to be an extension of the Barkhausen noise \cite{Uehara86} due to domain wall depinning. Several theories including self-organized criticality \cite{Bak88} or the random field Ising model \cite{Dahmen96} have been developed to describe them.

In other cases, the avalanche fields are reproducible and independent of the sample so that their origin can be considered to be more intrinsic, for example related to a nucleation field \cite{Lhotel08, Lhotel}, or to a resonant field in the case of quantum tunnelling of magnetization \cite{Paulsen95, Hernandez05}. In these cases, at certain values of magnetic field, the relaxation can become orders of magnitude  faster and the avalanche phenomenon is produced by a cascade of spin reversals associated with a self-heating of the sample. Indeed, at very low temperature, many materials have a very small specific  heat and poor thermal conductivity. Therefore when a spin flips, for one reason or another, the Zeeman energy $\Delta S \cdot H$ ($S$ is the spin value and $H$ is the local field) released can result in a large increase in temperature in the local environment of the flipped spin. If the diffusion of the heat to the rest of the sample and to the cold reservoir is slow enough, then the local heating will have enough time to activate the surrounding spins causing them to reverse. A  chain reaction can occur resulting in a fast reversal of the magnetization. A substantial increase in the sample temperature is often observed during the avalanche. In some cases, such as the single molecule magnet Mn$_{12}ac$, the avalanche propagation was compared to a magnetic deflagration \cite{Suzuki05, Macia07}. 

In \DyTi, when the field is along [111], the avalanches have been shown to disappear when the field sweeping-rate is very slow \cite{Fennell05,Slobinsky10}, suggesting a non-intrinsic behavior.
In the present paper, we have performed a systematic study of magnetic avalanches in \DyTi, measuring three different samples and applying the field along three directions. Our measurements show that the avalanche phenomenon does not depend on the direction of the magnetic field, despite the strong differences between the field induced equilibrium states in \DyTi . The three samples  qualitatively show the same features, but the fields at which the avalanches occur is sample dependent. We also performed a time-dependent study which shows that the avalanche propagation is very slow compared to previously reported avalanches. The underlying frustrated network, which prevents the magnetic excitations or monopoles from moving freely, in the lattice seems to be a clue to understand this behavior. 

\section{Samples and experimental details}

We have performed a systematic study of the avalanches using three different samples. For one sample we measured along three different axes, and for another along two axes, in order to understand the role of the field direction.
The three single crystals of Dy$_{2}$Ti$_{2}$O$_{7}$ were grown by the floating zone method as described in Ref. \onlinecite{Matsuhira02} (samples 1 and 2, grown at the Kyushu Institute of Technology) and Ref. \onlinecite {Prabhakaran11} (sample 3, grown at Oxford University). 

The samples were of dimensions shown in Table \ref{SampleTable} and were measured using low temperature superconducting quantum interference device SQUID magnetometers equipped with a miniature dilution refrigerator developed at the Institut N\'eel-CNRS Grenoble \cite{Paulsen01}. The samples were attached to a copper sample holder suspended from the dilution units mixing chamber which descends through the bore of the magnet.

The samples were aligned with the applied field parallel to their long axis to minimize the demagnetizing effects, that is to say along [001] for sample 1 and along [111] for samples 2 and 3. In addition, sample 1 was also measured along the  [111] direction and with the field along an arbitrary direction, and sample 2 was measured with the field perpendicular to the [111] axis. Finally, sample 2 was filed into an ellipsoid shape and then measured again along the [111] axis (sample 2* ellipsoid). The crystal alignment with respect to the field is accurate to within a few degrees.
The demagnetization factors $N$ (See Table~\ref{SampleTable}) were calculated with the analytical form for a rectangular prism \cite{Aharoni98}, except for the 2* ellipsoid which was estimated from Ref \onlinecite{Osborn45}.

\begin{table}[h]
\begin{tabular}{|*{5}{c|}}   \hline \hline
Sample& Dimensions (mm) & Mass (mg) & Direction & $N$ (cgs) \\ \hline
1 & 3.50 $\times$ 2.13 $\times$ 1.75 & 92.4 & [001] & 2.69 \\ 
1&"&"& arbitrary & 4.77 \\
2 & 3.80 $\times$ 1.85 $\times$ 0.90 & 44.2 & [111] & 1.74 \\ 
 2 &"&"& perp [111] & 3.65 \\ 
  2*ellipsoid & 3.74 $\times$ 1.84 $\times$ 0.76  & 25.1 & [111] & 1.18 \\ 
3 & 5.05 $\times$ 1.50 $\times$ 1.50 & 73.4 & [111] & 1.59 \\
\hline \hline
   \end{tabular}
\caption{Sample Details}
\label{SampleTable}
\end{table}

The avalanche measurements were performed using a magnetometer equipped with a solenoid capable of producing fields up to 3900 Oe with high resolution and a well defined zero field. 
The set-up can measure absolute values of the magnetization by the extraction method. To capture fast changing magnetization, a relative mode can also be used.

Two types of experiments were performed: magnetization as a function of field when the field is swept continuously (See Section \ref{MH}), and magnetization as a function of time after a given field is applied (See Section \ref{Mt}). 
For both measurements, the absolute values of the initial magnetization were first determined by the extraction method. Then the sample was positioned in one of the superconducting detection coils and measurements were made in the relative mode at about 10 points/second. The temperature was also measured continuously. After a certain time, depending on the type of measurement, 
and when the magnetization was no longer changing quickly, extraction measurements were again performed, and the relative measurements were then offset to fit the initial and later absolute value points.

\section{Avalanches of the magnetization}
\label{MH}

In an earlier paper reporting on details of avalanches in Dy$_{2}$Ti$_{2}$O$_{7}$, Slobinsky {\it et al.} \cite{Slobinsky10}  described many interesting features about the low temperature magnetic avalanches when the field is aligned along the [111] direction. They showed that; i) the avalanches can be suppressed by lowering the field sweeping rate; ii) the temperature rises in the sample during the avalanche; iii) the avalanche field (field at which the avalanche triggers) depends on the initial magnetization; iv) the magnetization at the end of the avalanche corresponds to the equilibrium magnetization curve at relatively high temperature (700 mK in their experiment). 
However, their experiment could not capture the time dependence of the avalanche. 

\begin{figure}[h]
\centering
\includegraphics[width=8cm]{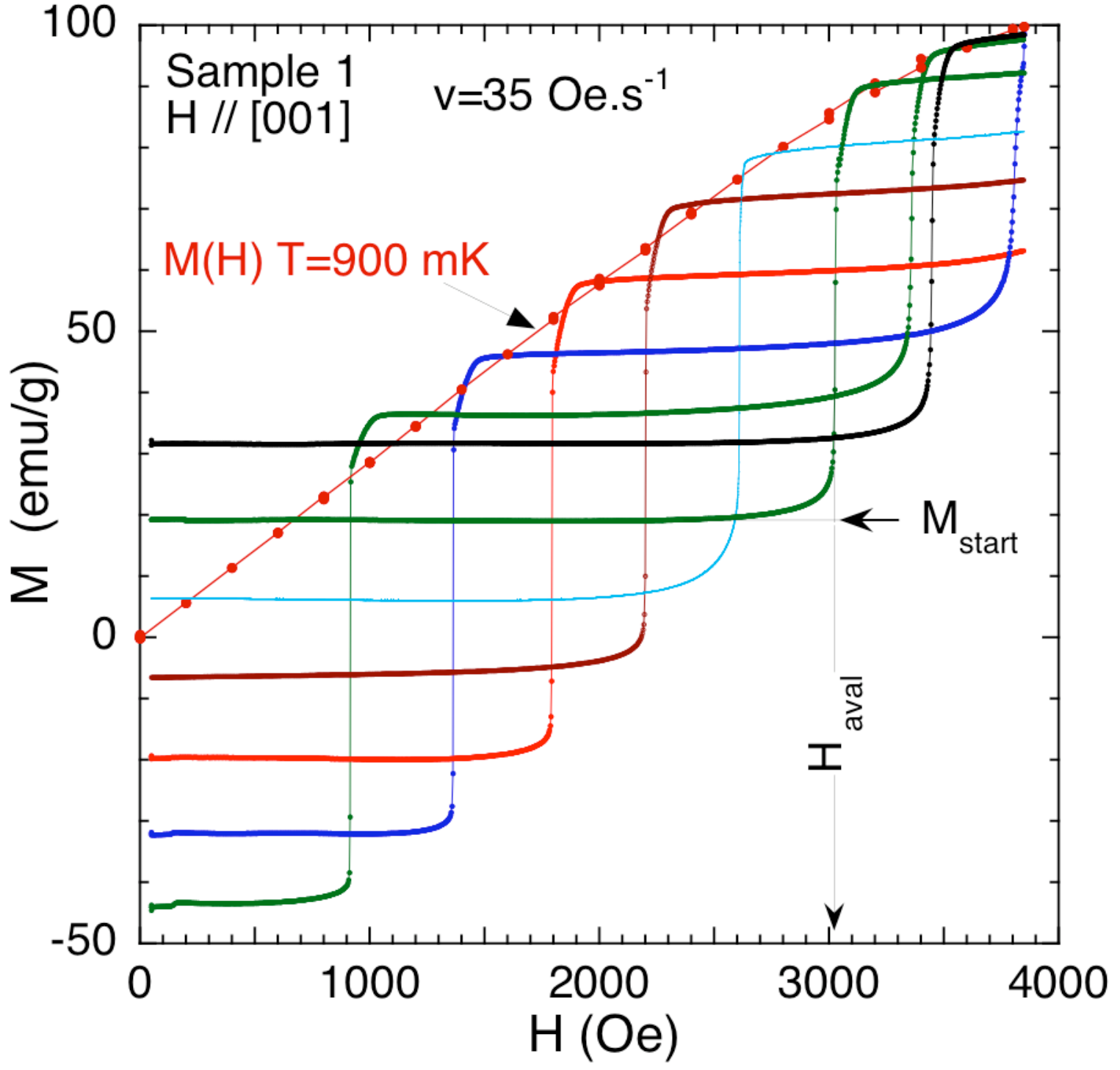}
\caption{(color online) Magnetization $M$ vs the applied field $H$ along the [001] direction for sample 1, starting from different field cooled states (from the bottom to the top: $H_{\rm FC}=-1400$ (green) to +1000 Oe (black) in steps of 400 Oe). The field sweeping rate was 35 Oe$\cdot$s$^{-1}$ (=0.21 T$\cdot$min$^{-1}$) and the starting temperature before the avalanche was 75 mK. Also shown is the isothermal $M$ vs $H$ measured at 900 mK (red circles). The avalanches are clearly seen as a sudden increase in magnetization at certain critical applied fields. An example of the definition of $M_{\rm start}$ and $H_{\rm aval}$ are shown for the 600 Oe field cooled curve.}
\label{fig_ava_001}
\end{figure}

In our experiments, avalanches appear below 500 mK, and are more pronounced when the temperature is lowered.  Furthermore, a relatively fast field sweeping rate is needed to trigger the avalanches: no magnetic avalanches were observed for field sweeping rates $v$ below 2.0 Oe$\cdot$s$^{-1}$. This field sweeping rate is somewhat slower than those of  Slobinsky \cite{Slobinsky10} (jumps for $v>0.025$ T.min$^{-1}= 4.14$ Oe$\cdot$s$^{-1}$) and Erfanifam \cite{Erfanifam11} (peaks at 0.015 T.min$^{-1} = 2.5$ Oe$\cdot$s$^{-1}$). These differences between experiments confirm that the avalanches are an out-of-equilibrium process which strongly depends on the thermal coupling between the sample and the cooling power of the experiment (mixing chamber of the dilution refrigerator or $^3$He chamber). Presumably,  the stronger the coupling, the higher the field ramping rate needed to trigger avalanches. 

\begin{figure}[h]
\centering
\includegraphics[width=8cm]{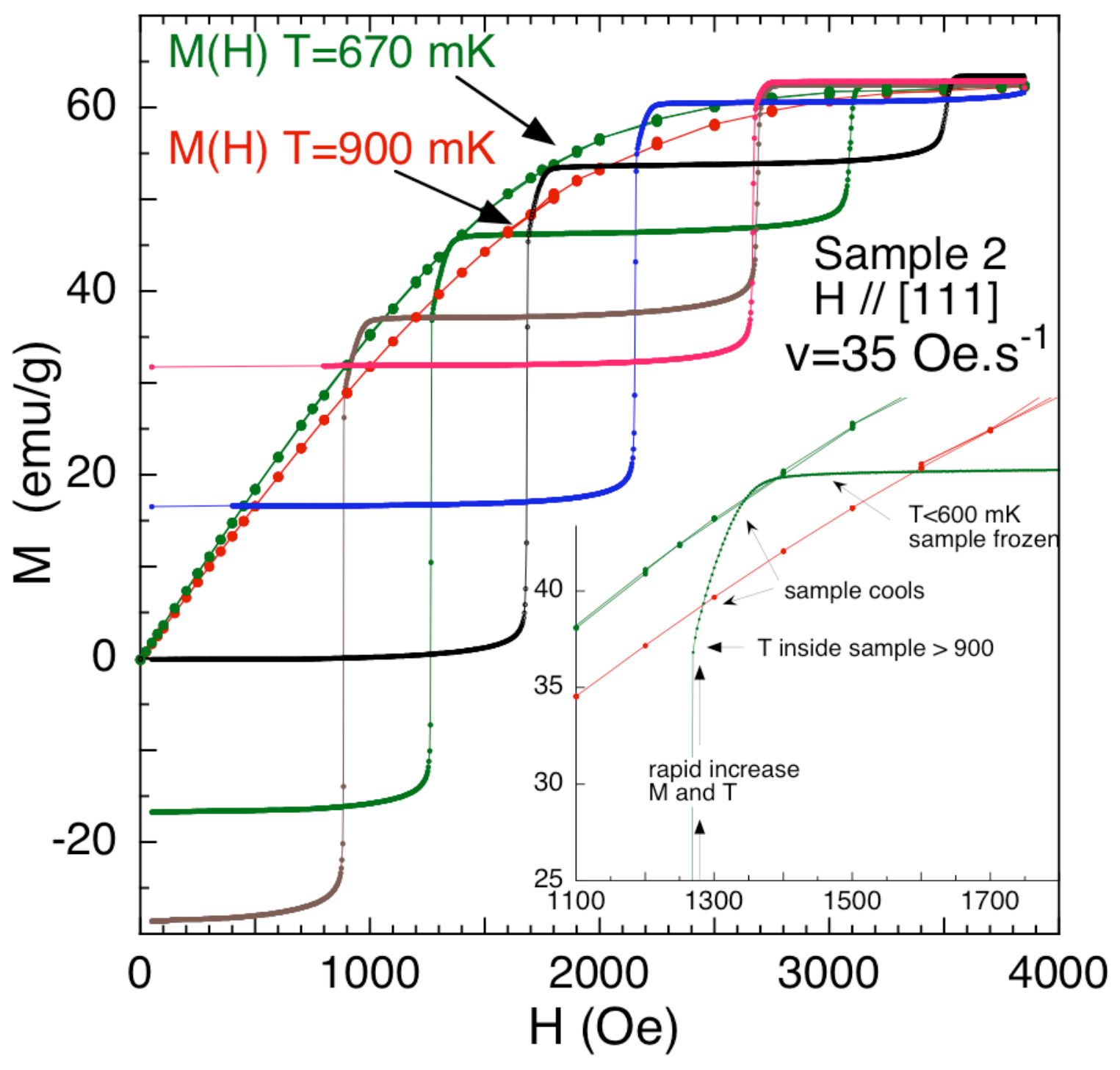}
\caption{(color online) Magnetization $M$ vs the applied field $H$ along the [111] direction for sample 2, starting from five different field cooled states  $H_{\rm FC}=-800$, -400, 0, 400 and 800 Oe (from the bottom brown, green, black, blue and red respectively). The field sweeping rate is 35 Oe$\cdot$s$^{-1}$ and the starting temperature before the avalanche was 75 mK. Also shown is the isothermal $M$ vs $H$ measured at 670 (green circles) and 900~mK (red circles).  The insert is a close up showing the end of an avalanche for the -400 Oe field cooled curve and shows that the magnetization increases very rapidly until it reaches its 900 mK equilibrium value, then it continues to increase but at a much slower rate, until falling out of equilibrium and becoming frozen when the sample temperature dips below approximately 600 mK. }
\label{fig_aval_900mK}
\end{figure}

For a given field sweep rate, a series of magnetization curves as a function of field were made for different values of the starting field cooled (FC) magnetization.  Some typical curves are shown in Figure \ref{fig_ava_001} for sample 1 with the field along the [001] direction and in Figure \ref{fig_aval_900mK} for sample 2 with the field along the [111] direction. The experimental protocol was the following: i) The sample was first heated to 900 mK for about 1 minute, and a field $H_{\rm FC}$ ranging from $-2000$ to 1000~Oe, was applied. ii) The heater power was cut, and the sample was rapidly cooled.  After a wait time of minutes,  the base temperature of 75 mK was achieved, and the absolute value of magnetization was measured to get the starting magnetization $M_{\rm start}$. iii) The sample was then moved to the center of one of the detection coils, and the field was set to 0 before being increased at a speed $v$. The relative magnetization was measured continuously at a sampling rate of 10 points/sec whilst the field is ramped. iv) At the end of the ramp (or sometimes just after an avalanche was detected), the field ramp was stopped and the absolute value of the magnetization was measured again by the extraction method, and the relative measurements were adjusted accordingly.

The different $H_{\rm FC}$ that were applied during cooling resulted in different starting magnetizations $M_{\rm start}$.  After field cooling, and at the beginning of the field sweep, the sample temperature is approximately 75 mK, and the magnetization was essentially frozen. It was observed that the magnetization remains more or less frozen at its starting value  (depending on the ramping rate) as the field is increased, until a critical field is approached, and then the magnetization suddenly jumps or ``avalanches''. The inset of figure \ref{fig_aval_900mK} shows that the magnetization increases rapidly until it reaches (approximately) its 900~mK equilibrium value, implying that the sample temperature is close to 900 mK, a value somewhat higher than that reported by Slobinsky \cite{Slobinsky10}.  Just after the avalanche, the magnetization continues to increase slightly, however it does so much more slowly, and in fact, the magnetization increases even if we stop the field ramp. 
This indicates that the sample temperature is rapidly decreasing after the avalanche, and at some point as it dips below 650 mK or so, the sample falls out of equilibrium, and the magnetization again becomes frozen on the time scale of the experiment.  The magnetization after an avalanche is more or less  sandwiched between the 900 and 670 mK equilibrium values. The magnetization remains at this new value on increasing the applied magnetic field until perhaps a second avalanche comes about. 

\begin{figure}[h]
\centering
\includegraphics[width=8cm]{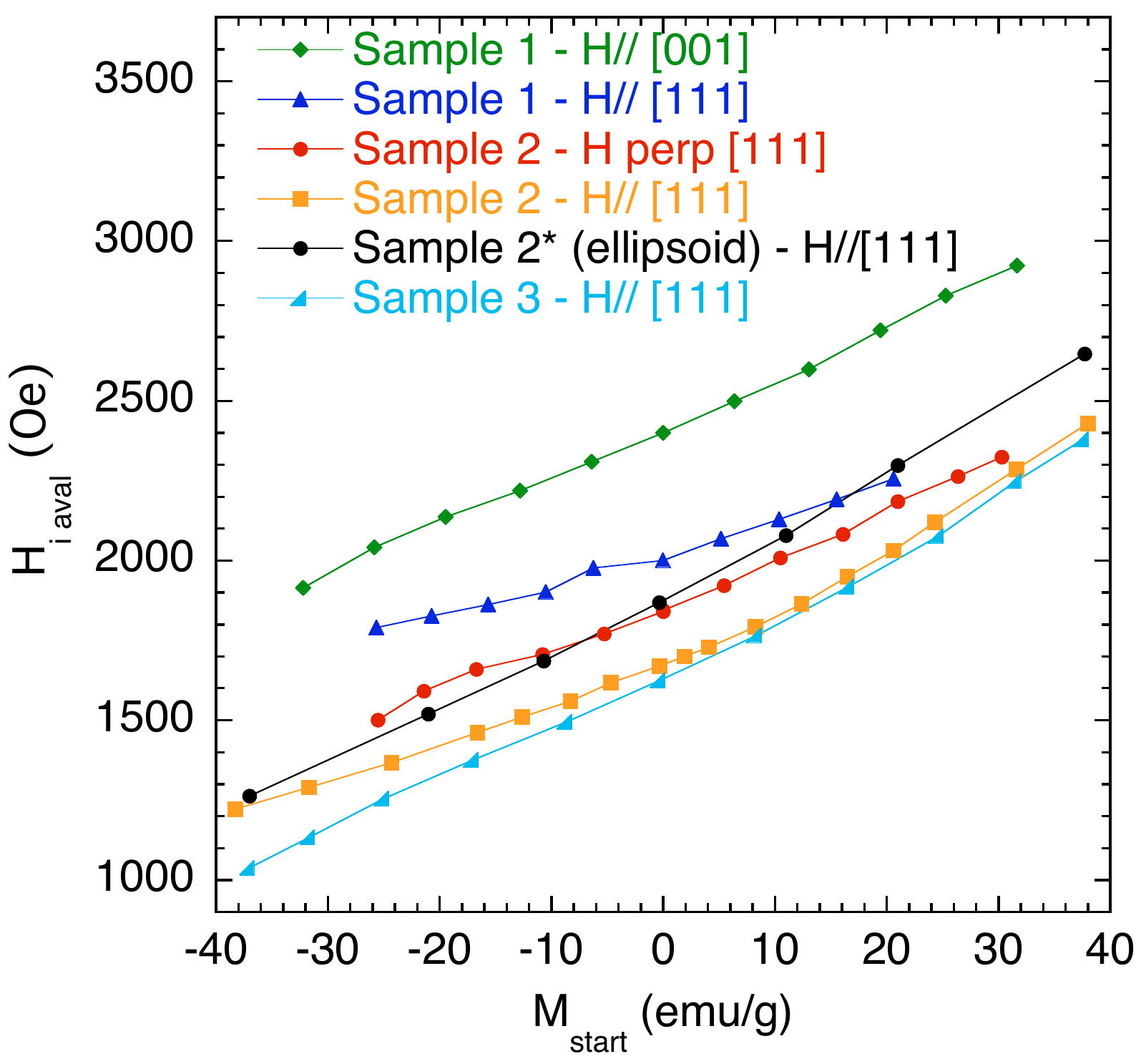}
\caption{(color online) Internal avalanche field $H_{\rm i aval}$ as a function of the initial magnetization $M_{\rm start}$ for sample 1 with the field along the [001] direction (green) and along [111] (blue), for sample 2 and 2*  along the [111] direction (orange and black respectively) and sample 2 perpendicular to [111] (red), and for sample 3 measured with the field along the [111] direction (light blue).}
\label{fig_AvalancheField}
\end{figure}

Qualitatively similar behavior was also observed when samples 1 and 2 were rotated as indicated in Table \ref{SampleTable}. However, the fields at which avalanches occurred were different, even after corrections for demagnetization effects. 
These results are summarized in Figure \ref{fig_AvalancheField}, where the demagnetization corrected internal field at the moment the avalanche begins, $H_{\rm i  aval}$,  is plotted as a function of the starting magnetization $M_{\rm start}$ for all three samples and the various orientations. We define the avalanche field using a tangent to the near vertical jump and its intercept with $M_{\rm start}$, an example of which is shown in Figure \ref{fig_ava_001}. The $H_{\rm i aval}$ dependence as a function of $M_{\rm start}$ is monotonically increasing and very roughly linear for all samples and field directions. But as can be seen, it also depends on the sample direction. The [111] direction seems to give lower avalanche fields for both samples 1 and 2. It is particularly interesting  that samples 2 and 3, although synthesized in two different laboratories \cite{Matsuhira02, Prabhakaran11}, show nearly the same dependence of the avalanche field on $M_{\rm start}$ whereas sample 1 along the [111] direction has much larger avalanche fields, even after corrections for demagnetization. 

The importance of the demagnetization correction and the associated limitations are also clearly shown in Figure \ref{fig_AvalancheField}. The ellipsoid sample, which is filed from sample 2, when compared has different internal avalanche fields. This demonstrates empirically the importance of the demagnetization fields, naively this discrepancy in the data can be ascribed to the fact that macroscopic demagnetization factors cannot accurately describe field fringing or focusing effects around corners. For example, a recent study~\cite{Bovo13} emphasized how in non-ellipsoidal samples of spin ice, the standard demagnetizing correction becomes a particularly poor approximation. There is even a possibility that the magnetic avalanches can nucleate in the corners of the sample. 

\begin{figure}[h]
\centering
\includegraphics[width=8cm]{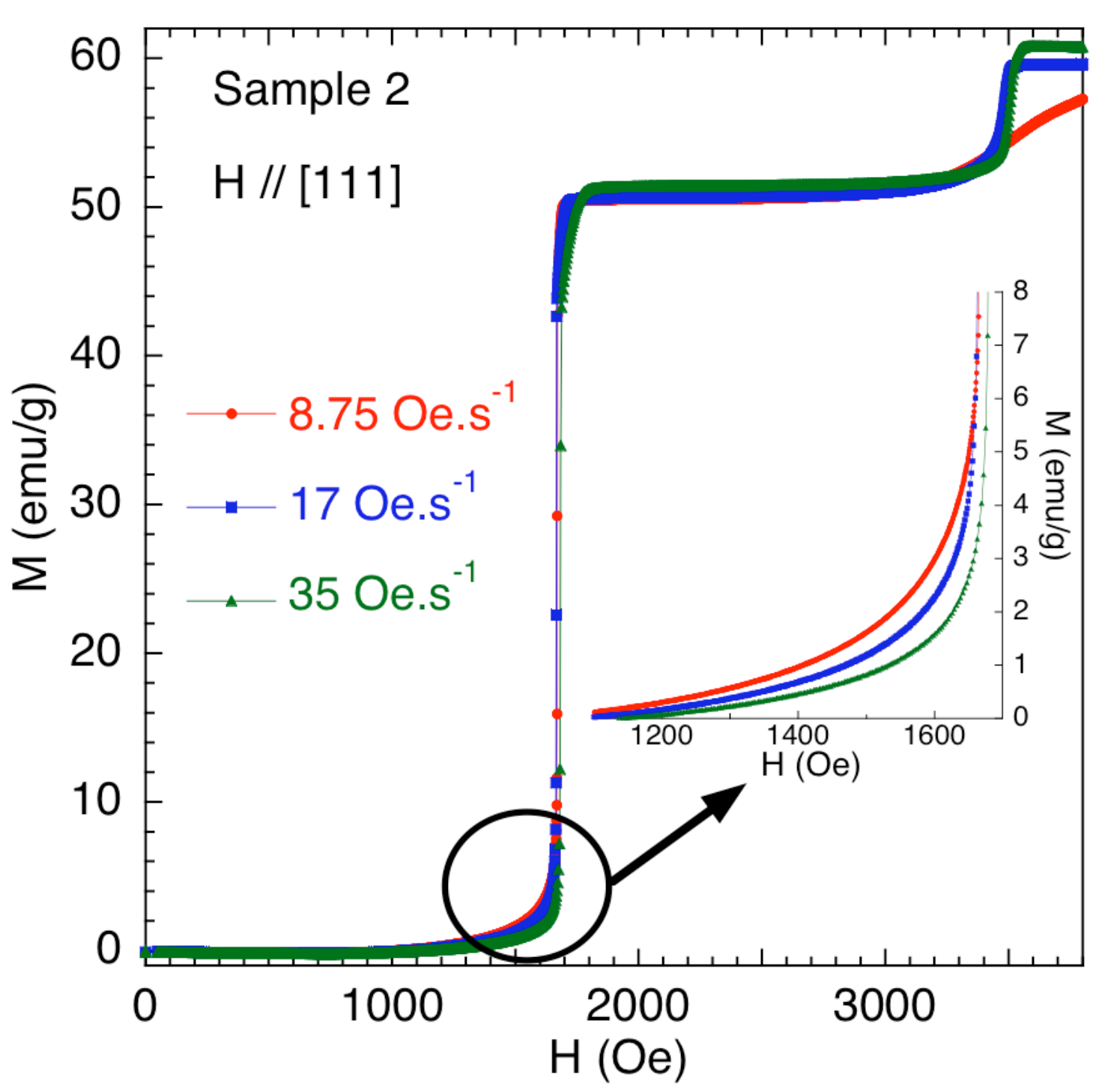}
\caption{(color online) Magnetization as a function of the applied field along the [111] direction at 82 mK for sample 2 at the various indicated field ramp rates, starting from a ZFC state. The field at which the avalanche occurs is nearly independent of the ramp rate. Inset: Closer inspection of the approach to the avalanche field.}
\label{fig_RateComparison}
\end{figure}

We have also observed that the avalanche field is nearly independent of the field ramping rates (in our measurement range from 8.75 to 70 Oe$\cdot$s$^{-1}$) (See Figure \ref{fig_RateComparison}). However, it can be seen in the inset of Figure \ref{fig_RateComparison} that the approach to the avalanche does depend on the ramping rate. A slower rate yields a more gradual approach since it gives the sample more time to relax. In fact, if the field ramp rate is too slow, less than 2 Oe$\cdot$s$^{-1}$, the avalanche becomes ``stretched out'' and more or less disappears, the sample has enough time to relax and dissipate the heat, and a more ``normal'', but distorted magnetization curve is observed. 

\begin{figure}[h]
\centering
\includegraphics[width=8cm]{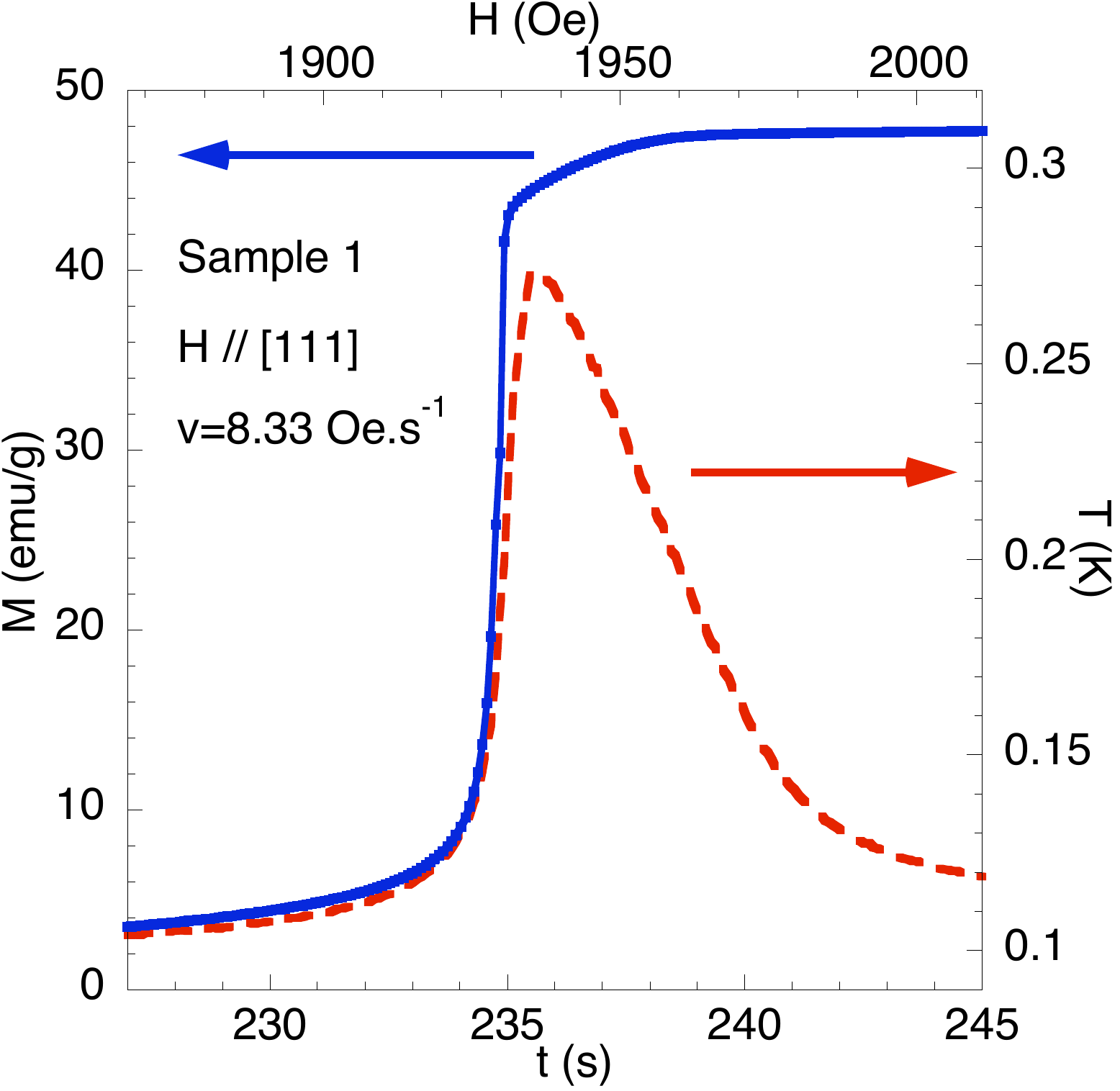}
\caption{(color online) $M$ (blue circles) and $T_{\rm thermo}$ (red dotted line) vs time $t$ during an avalanche for sample 1 with $H \parallel$ [111], $H_{\rm FC}=0$ Oe at 76 mK. The corresponding applied field $H$ is indicated on the top axis. $t=0$ matches with the starting of the field ramp. }
\label{fig_TempSpike}
\end{figure}

In Figure \ref{fig_TempSpike}, we show the temporal variation of the magnetization and the associated temperature spike during an avalanche. We confirm, as previously noted \cite{Slobinsky10, Erfanifam11}, that the avalanches are accompanied with an increase of temperature. The temperature maximum is of the order 250 mK. However, we must point out that the thermometer is located 20 cm above the sample and close to the mixing chamber, and therefore cannot indicate the instantaneous temperature spike inside the sample. We estimate experimentally that the sample temperature actually increases to approximately 900 mK, since the magnetization increases to the equilibrium value of this temperature.

\begin{figure}[h]
\centering
\includegraphics[width=8cm]{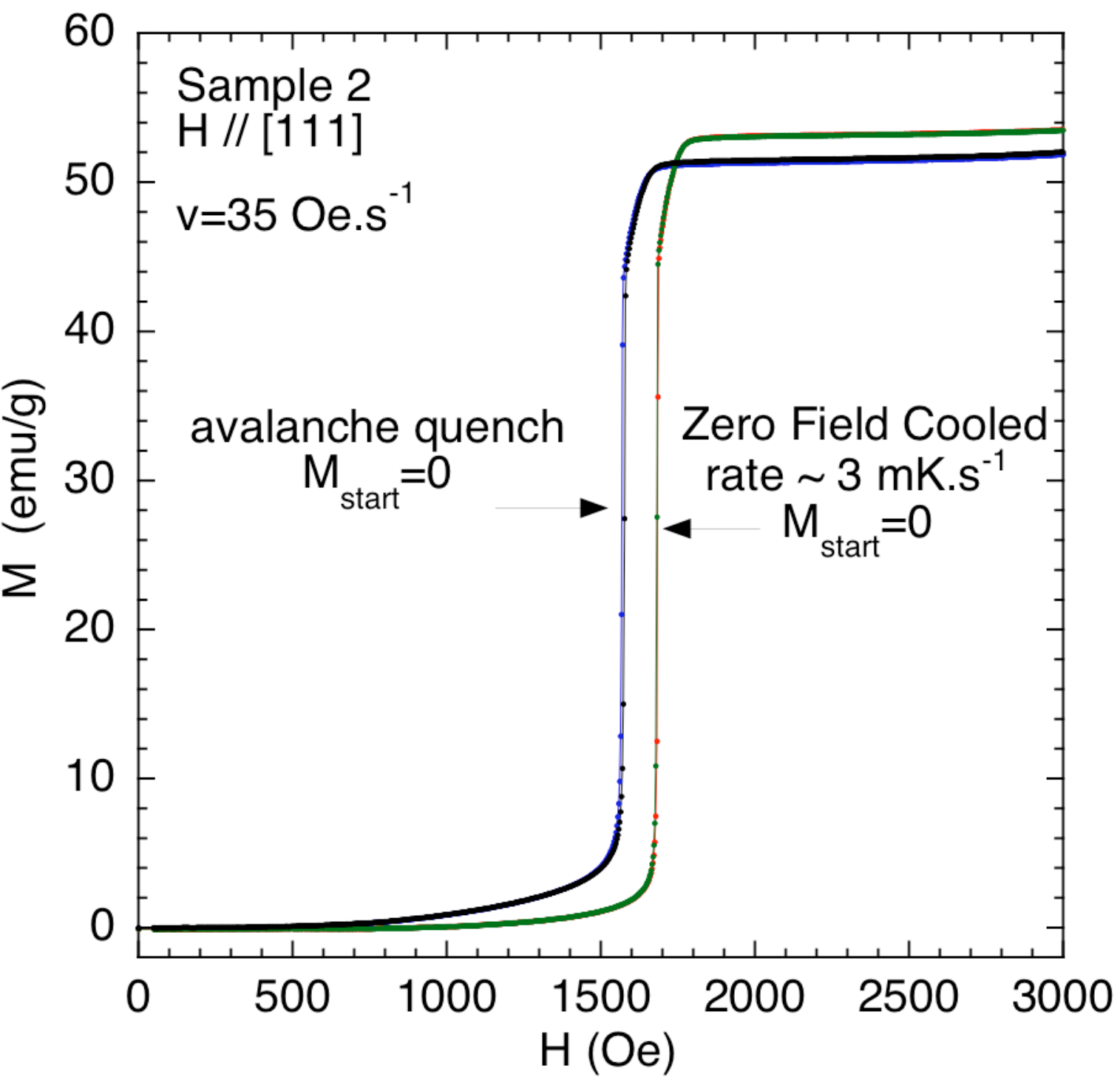}
\caption{(color online) Magnetization $M$ vs the applied field $H$ along the [111] direction for sample 2, where the initial starting magnetization $M_{\rm start}$=0 has been arrived at by using two different methods.  There are actually four curves, two for each method.  For the green and red curves $M_{\rm start}$=0 was attained by the conventional ZFC from 900 mK. For the blue and black curves the $M_{\rm start}$=0 was attained using the avalanche quenched protocol~\cite{ccp14}. } 
\label{M_start=0}
\end{figure}

On closer inspection of a given avalanche, we observe that this is actually a rather slow process, taking about 500 ms to 1 s for completion, as shown in Figure \ref{fig_TempSpike}. This corresponds to a propagation speed through the sample of a few millimeters per second. This value is very slow compared to avalanches in Mn$_{12}ac$ where the speed was estimated  \cite{Suzuki05} to be as fast as 10 m$\cdot$s$^{-1}$. Nevertheless, it is in the same range as the speed of the avalanches observed in molecular spin chains \cite{Lhotel}. 

Another curious effect is shown in Figure \ref{M_start=0}. There are actually 4 curves in the figure with $M_{\rm start}$=0, and were taken with the same starting temperature of approximately  75 mK, and with the same field ramping rate of 35 Oe$\cdot$s$^{-1}$. However, for one pair of curves, the sample was zero field cooled from 900 mK using the protocol described above before ramping the field. As can be seen, the two curves are nearly identical, falling one on top of the other, and both with an avalanche field at 1680 Oe. 

The other pair of curves are also nearly identical but they clearly differ from the first pair. For this pair of curves, the way we attained the $M_{\rm start}$=0 starting point was somewhat different.  The ``magneto-thermal avalanche quench'' technique was used to quickly cool the sample as described in detail in Ref.~\onlinecite{ccp14}, here by quickly switching ($<$ 0.5 s) the field from -3900 to +3900 Oe the magnetic Zeeman energy cannot escape the sample quickly enough so the sample heats to $\sim$ 900 mK, and then by again switching the field from +3900 to 0 Oe the sample can be heated and quickly zero field cooled from 900 mK. Only the sample is heated,  the surrounding copper sample and the mixing chamber remain at a very low temperature. The ensuing thermal quench of the sample is as fast as possible, and as a consequence, a much larger density of monopoles is frozen into the sample. We then waited for 10 minutes to make sure that the sample was at base temperature and to be consistent with the FC protocol, before starting the field ramp.  These two curves are quite different from the former: there is more relaxation approaching the avalanche field, and more obviously, the avalanche field is not the same as the ZFC curves, it is less, at approximately 1570 Oe. 

\begin{figure}[h]
\centering
\includegraphics[width=8cm]{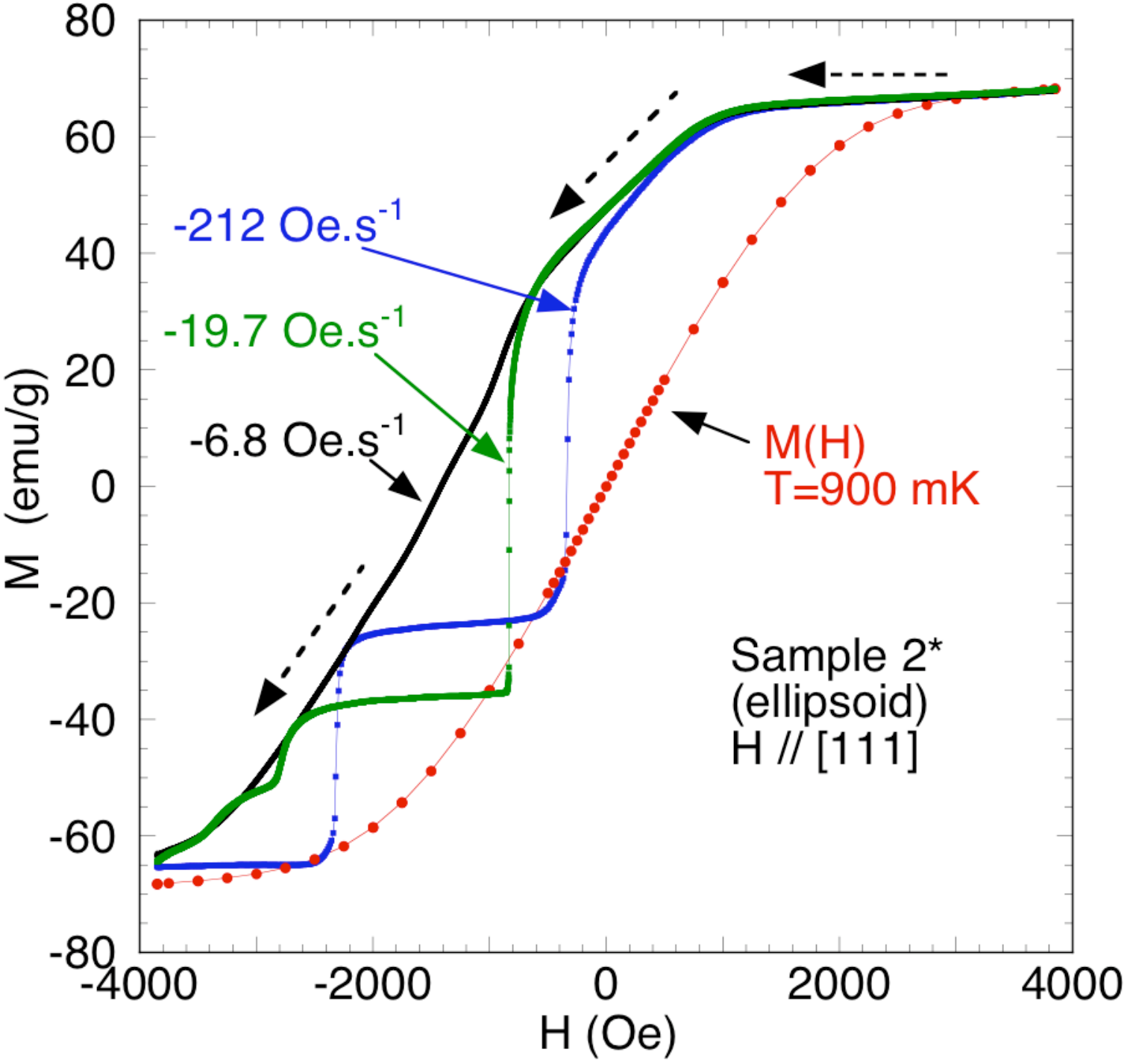}
\caption{(color online) Magnetization $M$ vs the applied field $H$ along the [111] direction for sample 2*, (ellipsoid) starting at + 3900 Oe for three different field ramping rates. The sample was first avalanche quenched from 0 to +3900 Oe, and then allowed to cool so that starting temperature before ramping was below 110~mK.  Also shown (red) is the equilibrium M(H) vs H measured at 900 mK. }
\label{fig_neg_ramp_1}
\end{figure}

Until now we have presented data with avalanches that start within the spin ice phase, and for the most part remain in the spin ice phase, or for larger fields cross over to the kagom\'e ice state. It is also interesting to look at avalanches that start from within the kagom\'e ice state and cross over to the spin ice phase \cite{mostame}. Examples of this are shown in Figure \ref{fig_neg_ramp_1}, the sample was first prepared by using the avalanche quench protocol by rapidly switching the field from 0 to 3900 Oe, followed by a wait period of 60 seconds for the sample temperature to drop below 110~mK. 

Figure \ref{fig_neg_ramp_1}  shows three examples with  different field ramp rates all starting from 3900 Oe. For the slowest field ramp of 6.8 Oe$\cdot$s$^{-1}$ the magnetization as a function of field is smooth, no avalanches take place and no spikes in the temperature could be detected. However for faster ramp rates avalanches were observed, two examples for 20 and 200 Oe$\cdot$s$^{-1}$ are shown.  In both cases the avalanches were complete, reaching the 900 mK thermal equilibrium magnetization value, but both show an important relaxation of M during the field ramps before the avalanche. More interesting is the very large difference in avalanche field as a function of ramping rate, which correlates with the very different values of  $M_{\rm start}$ for the two curves. These curves are in stark contrast to Figure \ref {fig_RateComparison}  for avalanches within the spin ice phase were the magnetization remains more or less constant and the avalanches occur at the same field for different field ramp rates.

\section{Relaxation of the magnetization}
\label{Mt}
The behavior of the avalanches is characteristic of an out of equilibrium process, mainly due to the difficulty for the materials at low temperature to dissipate the Zeeman energy which is created during spin reversals. In that sense, avalanches in \DyTi\ are quite classical and \DyTi\ behaves
like other magnetic materials which exhibit avalanches at low temperature. The avalanche field value $H_{\rm i aval}$  depends on the sample, on the field direction and on the initial magnetization (and how this was obtained), but is qualitatively robust against these parameters and very reproducible in given experimental conditions as Figure \ref{M_start=0} attests. However, a key difference between avalanches in \DyTi\ and others known magnetic avalanches is the slow propagation speed, and its long approach time so that it is very easy to measure the magnetization during the avalanche.  To further probe this time dependence, we have measured the relaxation of the magnetization starting from a ZFC state at different temperatures for various given applied fields. A curious result is that, as was suggested in the magnetization curves, the avalanche phenomena can occur after a time delay which can be larger than 1 s. 

\begin{figure}[h]
\includegraphics[keepaspectratio=true, width=8cm]{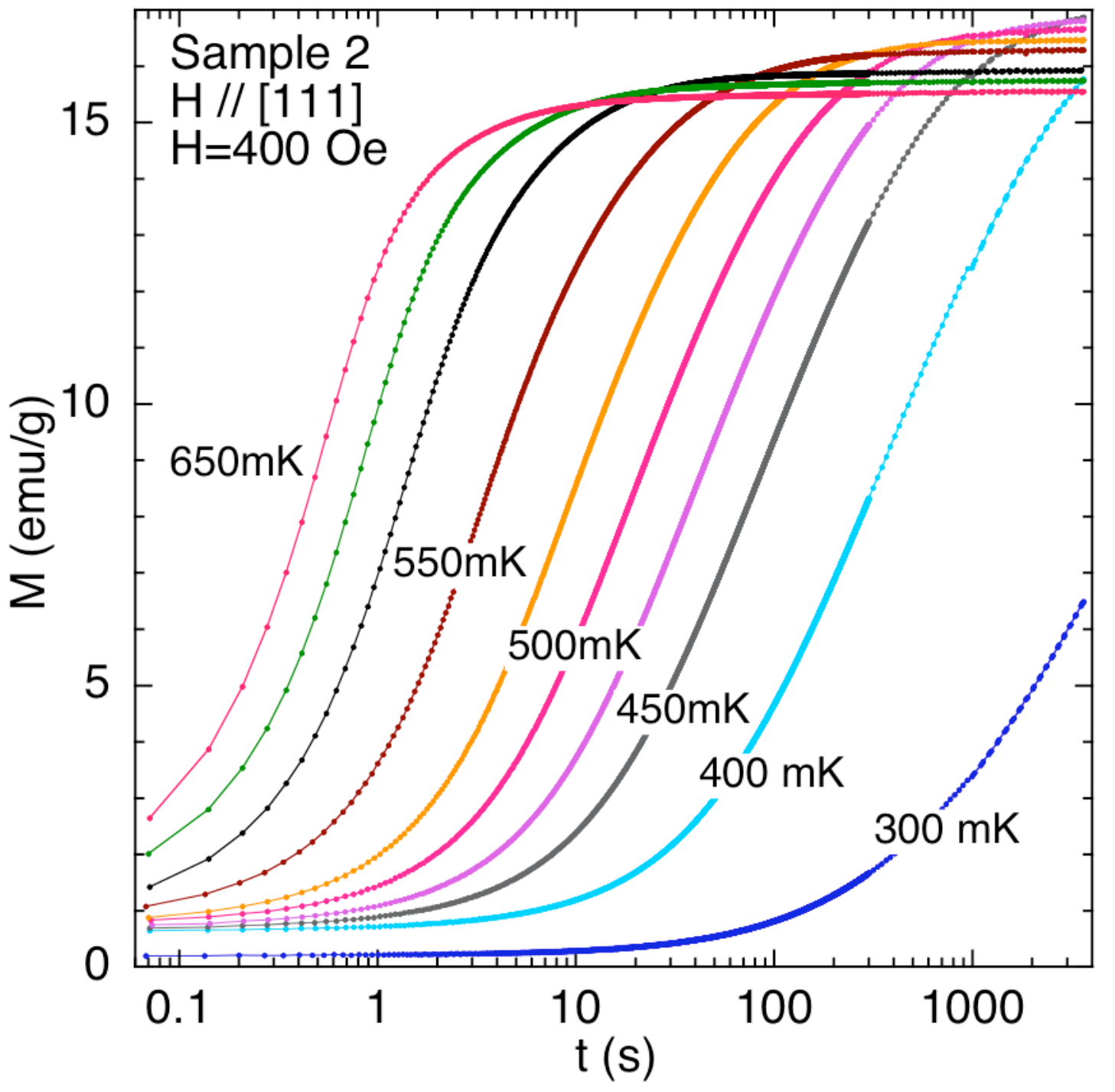}
\caption{(coor online) Relaxation of the magnetization $M$ vs time $t$, in a semilogarithmic scale, after ZFC, in an applied field of 400 Oe, at temperatures between 300 and 650 mK for sample 2. The field was along the [111] direction. Over this temperature range and for fields below 600 Oe the  relaxation is more or less normal, in the sense that it is roughly exponential, and there was no self heating of the sample, and no avalanche.}
\label{fig400G}  
\end{figure}

The experimental protocol for these ZFC relaxation measurements is the following:
i) before each measurement, the sample was first heated to 900~mK in zero field for 60 s;
ii) the heater power was then cut, and the sample was cooled to the lowest temperature (below 75 mK). The cooling rate was not constant, but was approximately $|dT/dt|=10$ mK$\cdot$s$^{-1}$ at 500 mK;
iii) there was a 10 min wait at the lowest temperature; 
iv) the temperature was then regulated to the target temperature, waiting for 5 min for thermalization;
v) the field was then applied, the timer set to zero and the relaxation of the magnetization was recorded in the relative mode. Then after 5 min, measurements were made using the extraction method for absolute value of the magnetization, and the relative measurements were adjusted.

The measurements presented in this section were made along the [111] direction. However, we  stress that the same qualitative features were observed in all the samples, and for whatever the field direction. 

Figure \ref{fig400G} shows characteristic relaxation curves taken  at constant  temperatures between 300 and 650~mK and in an applied field of 400 Oe for sample 2.  Over this temperature range and for fields smaller than 600 Oe, no self heating was detected by the thermometer, and no avalanches occur. The relaxation behavior in this ``low field limit''  is actually not simple \cite{Matsuhira11}, and will not be discussed here. However, suffice to say that for temperatures above 400 mK and when corrections are made for demagnetizing effects, the relaxation can be considered as exponential to a first approximation.

\begin{figure}[h]
\includegraphics[keepaspectratio=true, width=8cm]{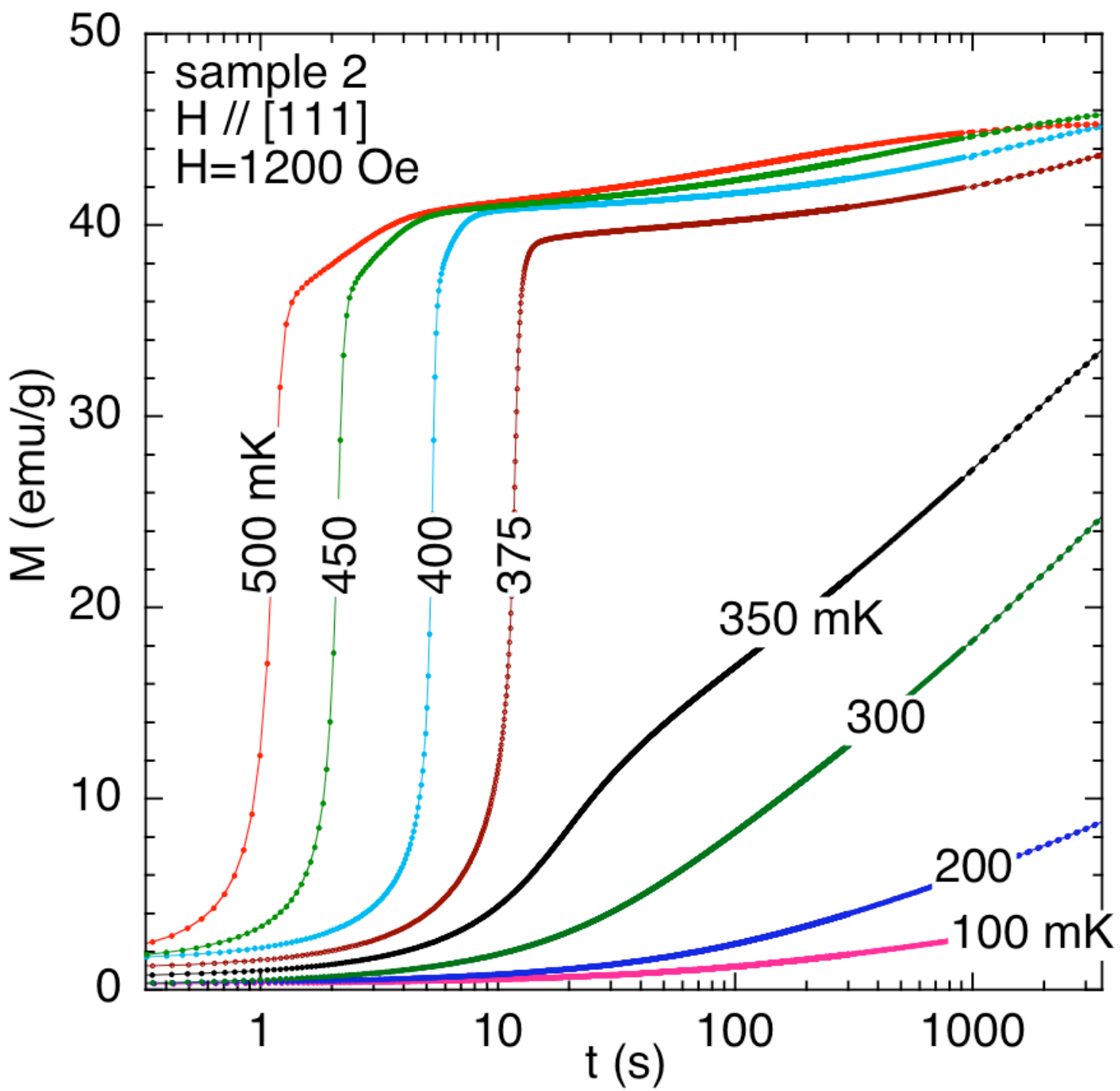}
\caption{(color online) Magnetization $M$ vs time $t$ in a semilogarithmic scale, after ZFC, in an applied field of 1200 Oe  for sample 2.}
\label{rlx_1200g}  
\end{figure}

\begin{figure}[h]
\includegraphics[keepaspectratio=true, width=8cm]{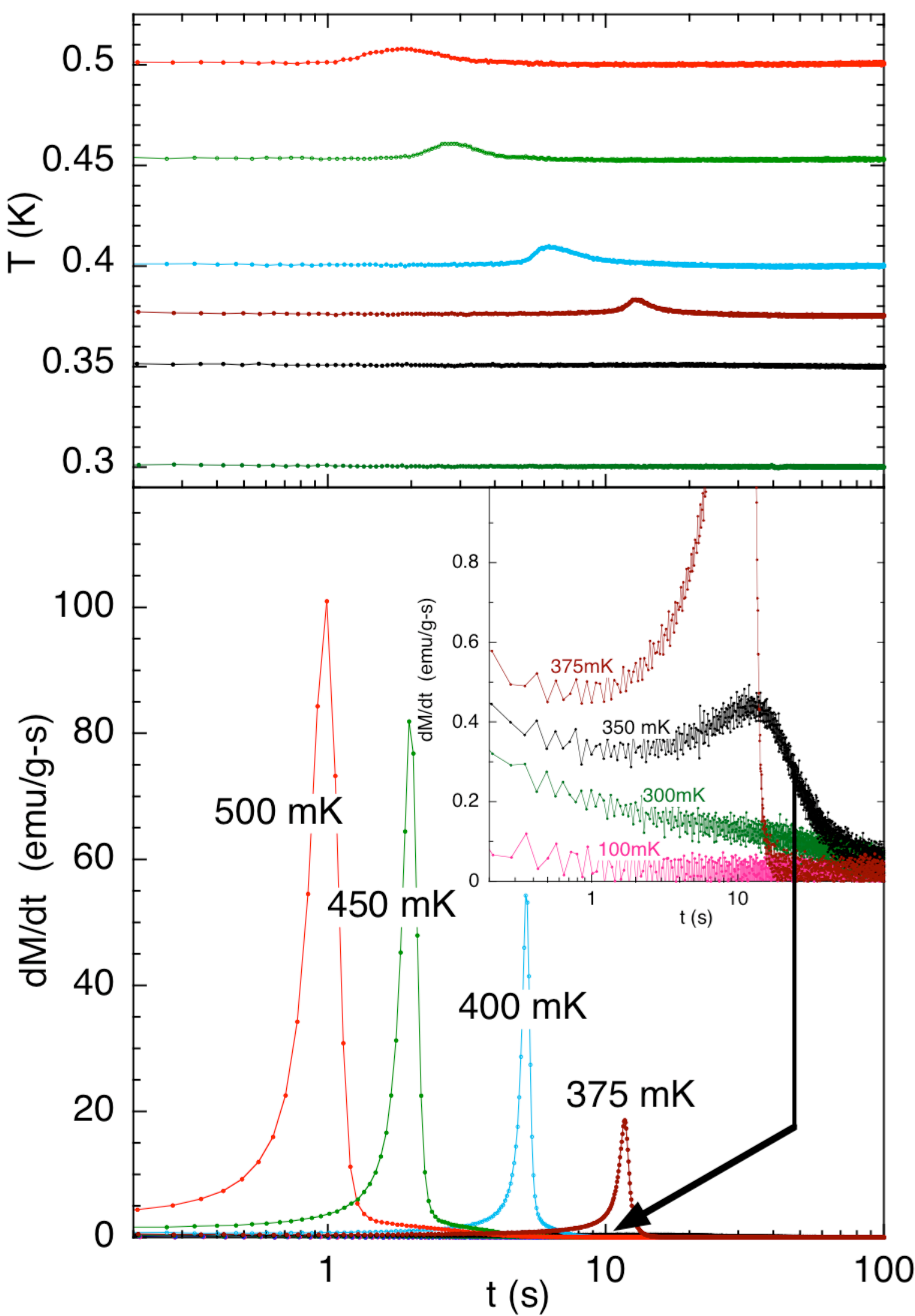}
\caption{(color online) Bottom:  $dM/dt$ vs time $t$ in a semilogarithmic scale, after ZFC, in an applied field of 1200 Oe for the data shown in Figure \ref{rlx_1200g}. The inset shows a closeup of the low temperature data showing a small peak for the 350 mK data, and below this temperature, no other peaks could be discerned. Top: The temperature during the relaxation measured on a thermometer 20 cm above the sample. The peaks indicate self heating of the sample during rapid magnetization changes.}
\label{dmdt_1200g}  
\end{figure}

When the relaxation is measured in fields greater than 600 Oe, and depending on the temperature, the shape of the relaxation curves can become distorted. For example, at 1200 Oe (See Figure \ref{rlx_1200g}), an anomalous shape of the relaxation curve is observed for temperatures $\ge$ 350 mK. In parallel, a spike in the temperature was  also detected at the thermometer for these curves as shown in the top panel of Figure \ref{dmdt_1200g}. Thus, the fast relaxation measured at 1200 Oe for $T\ge350$ mK are the analog of the magnetic avalanches reported in hysteresis loops in Section~\ref{MH}. On the other hand, below 300 mK,  no increase in the temperature of the sample could be detected, and the relaxation curves appear to be normal. 

Also shown in Figure \ref{dmdt_1200g}, is the derivative of $M$ with respect to time. For temperatures greater than 350 mK, $dM/dt$ has large, sharp peaks. This indicates that the relaxation of the magnetization is becoming faster with elapsed time for these curves. These peaks in $dM/dt$ are accompanied by a peak in the temperature. Although it is measured with a short delay in the top of Figure \ref{dmdt_1200g}, accurate measurements show that the temperature peak coincides with the $dM/dt$ peak.   
Note that even for 350 mK, a small maximum in $dM/dt$ can be seen in the inset, and a very weak peak is seen on the thermometer. In that sense, for an applied field of 1200 Oe, 350 mK is on the border of avalanche or normal relaxation.

We can break the avalanche process into three regimes: 1)  during the first pre-avalanche regime the magnetization slowly increases, and heat coming from the Zeeman energy released by the flipping spins causes the temperature to rise inside the sample which in turn decreases the relaxation time, causing more spins to flip and so on; 2) the second regime which takes about 1 second or so, the avalanche occurs because at some point the ever increasing Zeeman energy into the sample has overwhelmed the rate at which the sample can dissipate the heat, and as a result the temperature rises abruptly inside the sample; 3)  in the third regime, no more heat enters the sample, and the sample recovers and the temperature decreases, thus leading to a relaxation which is slowing down with time, before reaching the equilibrium value. These curves emphasize the originality of avalanches in spin-ice: i) they start after a quite long delay, from 1 to 10 seconds; ii) and they propagate slowly, in about 1 s. These features are most likely due to the topology of the spin-ice state, which impose strong constraints in the propagation of the spin-reversals or in the monopole velocity. 

\begin{figure}[h]
\includegraphics[keepaspectratio=true, width=8cm]{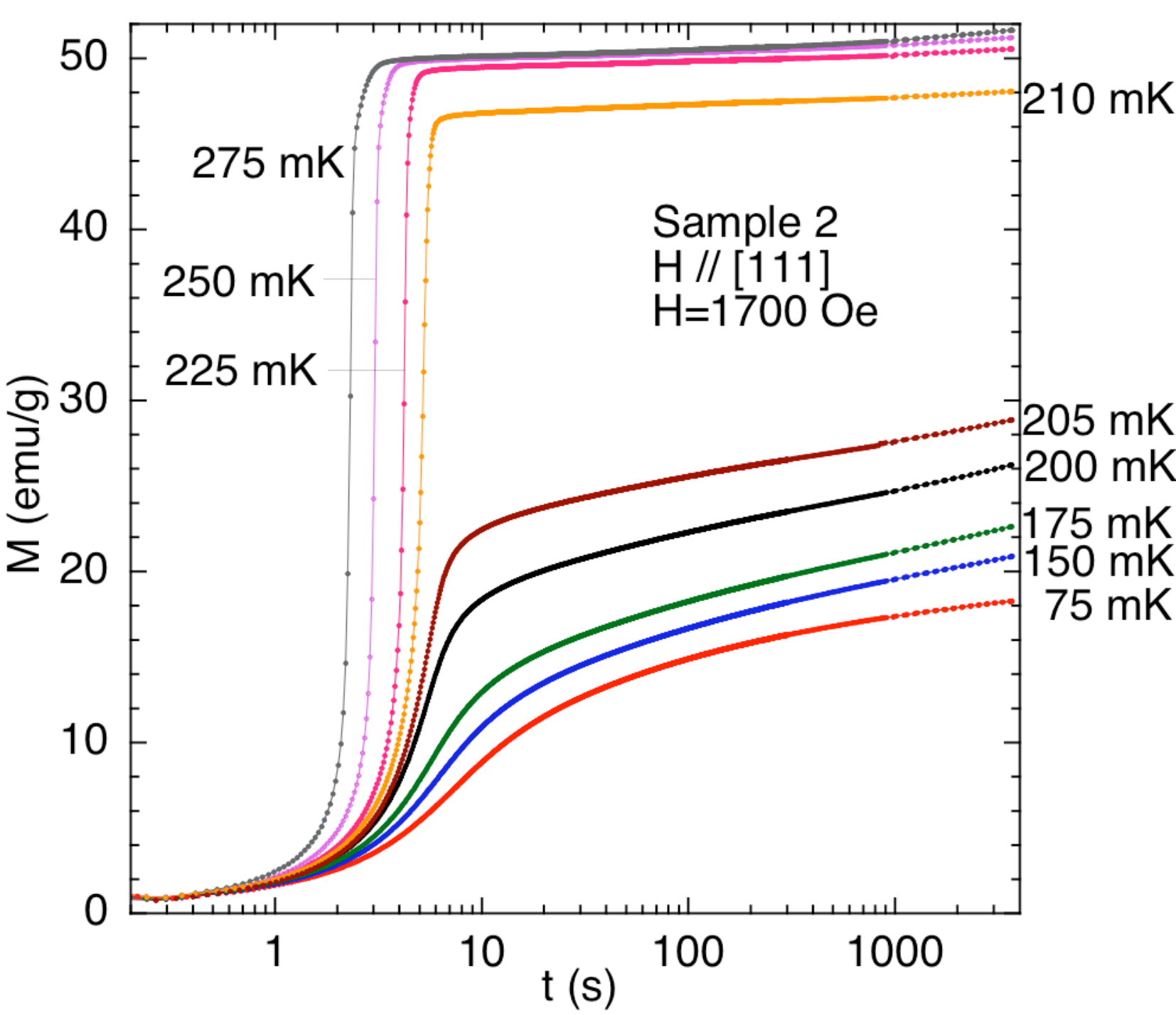}
\caption{(color online) Magnetization $M$ vs time $t$ in a semilogarithmic scale, after ZFC, in an applied field of 1700 Oe, at temperatures between 75 and 210 mK  for sample 2.}
\label{fig1700G}  
\end{figure}

For high enough fields,  avalanches took place even at the lowest temperature measured as  shown in Figure \ref{fig1700G} for relaxation in 1700 Oe. In this example, complete avalanches occurred for temperatures above 225 mK. However for lower temperatures, the avalanches were incomplete but nevertheless sizable temperatures spikes were recorded even at 75 mK indicating self heating for these data. When measuring in 2000 Oe, the avalanches were always complete, and the magnetization reaches its 900 mK equilibrium value just after.

\begin{figure}
\includegraphics[keepaspectratio=true, width=7.5cm]{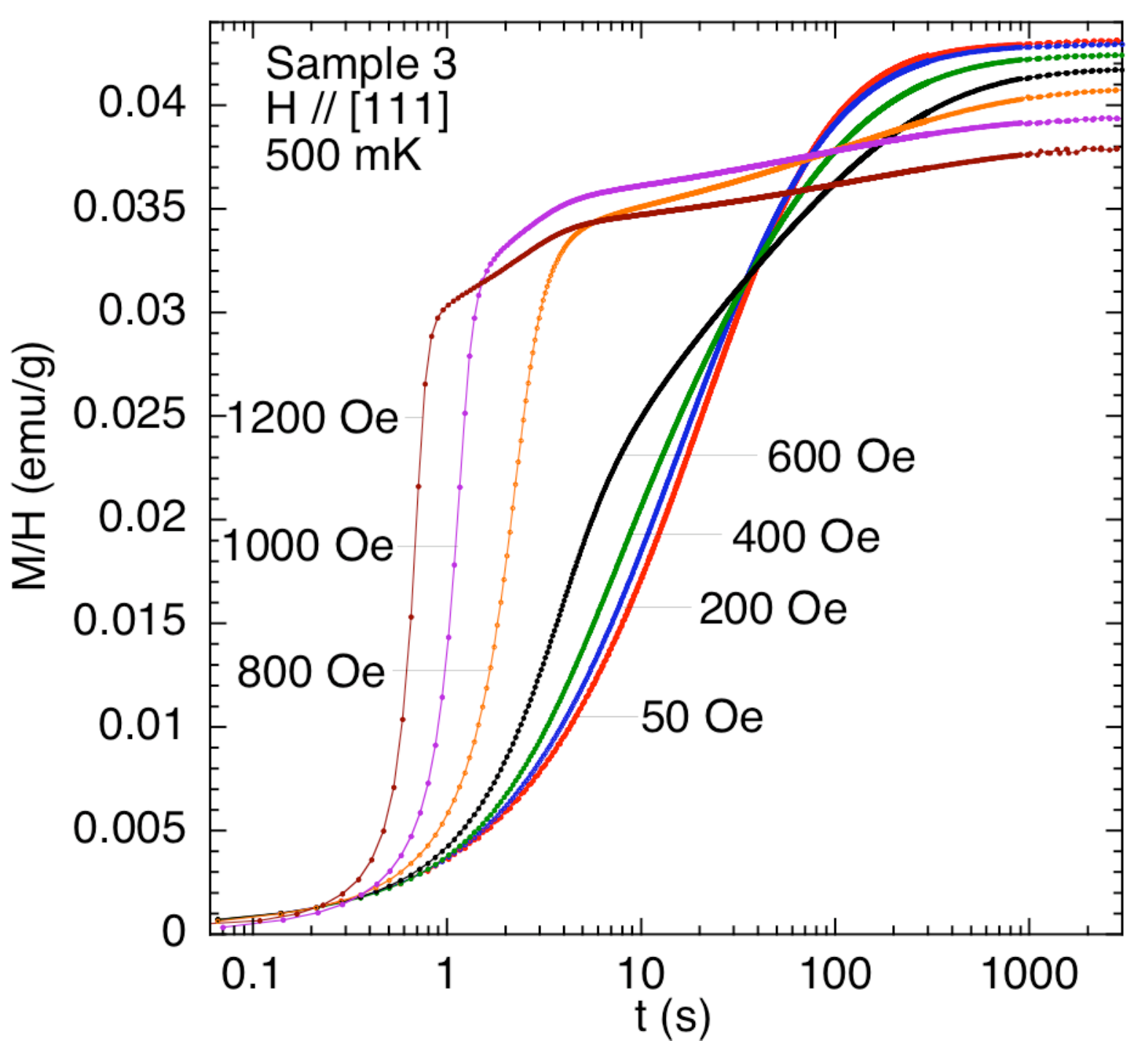}
\caption{(color online) $M/H$ vs time $t$ in a semilogarithmic scale, after ZFC, at 500 mK, for fields $H$ between 50 and 1200 Oe,  for sample~3.}
\label{fig500mK}  
\end{figure}

In Figure \ref{fig500mK},  the relaxation curves normalized by field $M/H$ measured at 500 mK for sample 3 for fields between 50 and 1200 Oe are plotted vs time. The same general features are observed as in sample 2, although the field range at which the avalanches occur is slightly shifted. Nevertheless, the same regimes are observed: slow relaxation at small fields, partial avalanches at intermediate fields, and fast relaxation associated with heating at larger fields. This figure also shows that for low fields ($H<200$ Oe), the relaxation curves almost superimpose one on top of the other, indicating that relaxation times are very weakly dependent on field at 500 mK. However we note that for $T< 300$ mK, the field dependence becomes increasingly important.

\begin{figure}
\includegraphics[keepaspectratio=true, width=8cm]{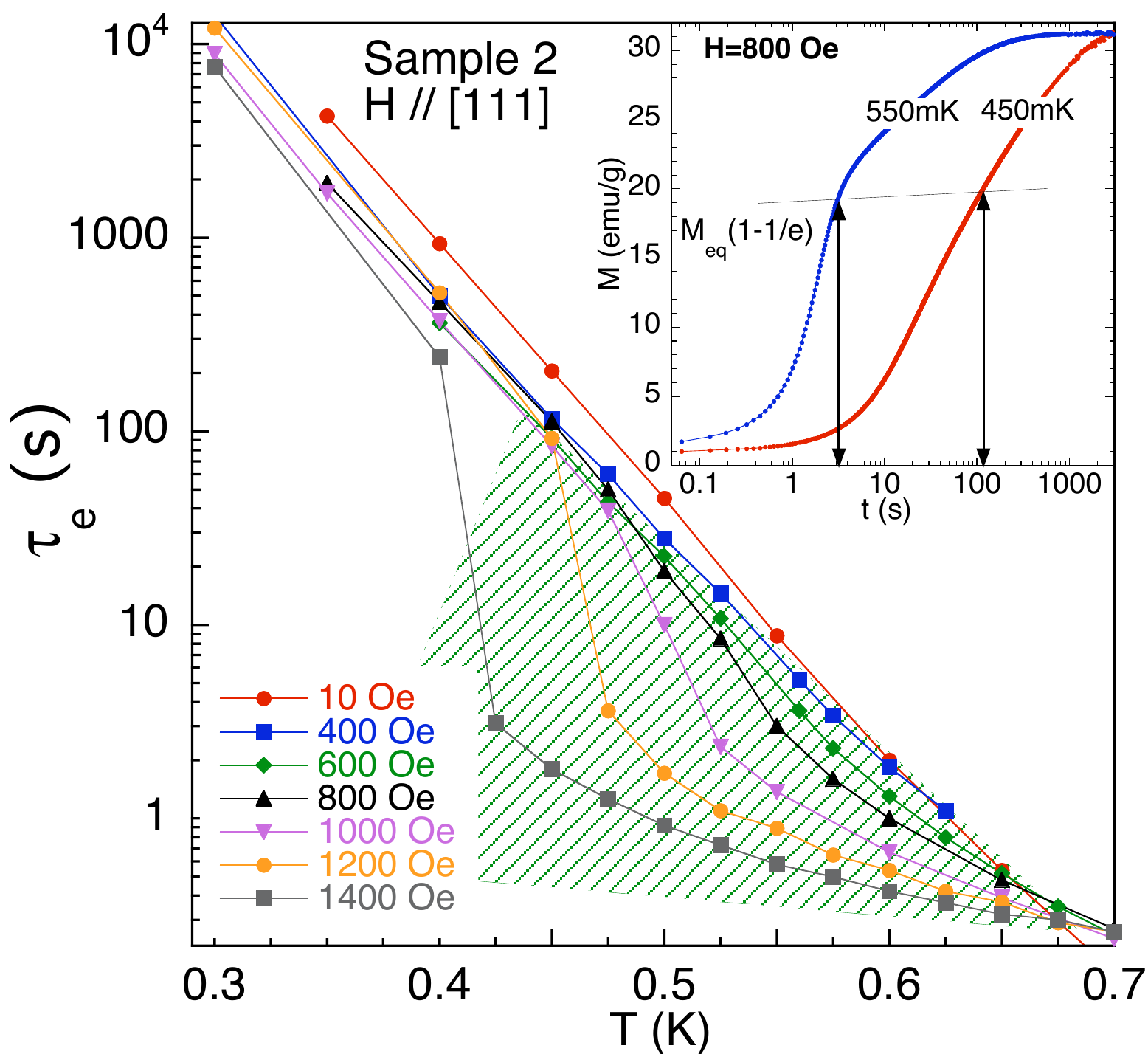}
\caption{(color online) The effective relaxation time $\tau$ vs temperature $T$ in a semilogarithmic scale for applied fields between 10 and 1400 Oe, for sample 2. The lines are just guides for the eyes. The green shaded region shows the zone in which the sample is self heating, resulting in a reduction of the relaxation time. Inset: Relaxation curves $M$ vs $t$ at 800 Oe showing the way of defining $\tau$ (see text). }
\label{figtau}  
\end{figure}

We summarize these observations by plotting the effective relaxation time as a function of temperature at different fields (See Figure \ref{figtau}). We define an arbitrary criterion to determine the characteristic relaxation time $\tau$: $\tau$ is the time at which the magnetization reaches the value $M_{\rm eq} (1-1/e)$, where $M_{\rm eq}$ is the final equilibrium magnetization (See inset of Figure \ref{figtau}). When the relaxation is normal, i.e. without avalanching, $\tau$ defined in this way is a reasonable quantity that characterizes the bulk of the relaxation. But when an avalanche occurs, this $\tau$ has no clear meaning because the temperature is not constant and the curves are far from being exponential. 

As shown in Figure \ref{figtau} for fields less than 600 Oe, when the relaxation occurs in thermal equilibrium without avalanches, all the $\tau$ vs $T$ curves collapse, and show only a weak dependence on the magnetic field in this temperature range. 
This is in contrast to the case for larger fields where depending on the starting temperature, the effective relaxation is very fast, and the points deviate strongly from the low field thermal equilibrium trend. Clearly local heating occurs which accelerates the relaxation in this region. However for a given (not too high) applied field, at lower temperatures  the effective $\tau$ can rejoin the thermal equilibrium line, indicating that at under these conditions avalanches are not taking place.  Thus there is a  regime sketched by the shaded region in Figure \ref{figtau} that delimits where avalanches take place and $\tau$ cannot be defined. Finally for fields above 1500 Oe, no reliable measurement of  $\tau$ can be made.

\section{Discussion and Conclusion}
\label{discussion}

The behavior of the avalanches in \DyTi~with respect to other materials is novel, and to explicitly understand the properties is likely complex, however it is instructive to highlight some fundamental properties. All the experimental data \cite{Slobinsky10}, including this work demonstrates under a constant field sweep the temperature of the sample increases to $\sim$ 700 to 900 mK, well above the blocking temperature often sited as $\sim$ 600 mK. 
This feature can be understood simply by considering the equilibrium properties of \DyTi.  As initially measured by Ramirez {\it et al.} \cite{Ramirez99}, the specific heat of \DyTi~increases strongly above 600~mK and presents a broad peak between 800 mK and 1.2~K. Considering roughly that the released energy during an avalanche is absorbed by the sample, $\Delta M \cdot H=C \Delta T$, the sample temperature will increase as long as the specific heat is low enough, and will saturate for a large specific heat: the obtained temperature 900 mK (See Figure \ref{fig_aval_900mK}) at the end of the avalanche is thus consistent with the known thermal properties of \DyTi. 

To further describe the avalanches mechanism, we however need to consider the out-of -equilibrium process and the heat flows during the avalanches. 
Indeed, for an avalanche to occur the Zeeman energy of flipping spins flowing into the sample overwhelms the energy absorbed and more importantly, the energy leaking out of the sample due to thermal conduction to the heat bath. The balance between these rates can become unstable due to a positive feedback; when the energy flowing in increases, it will heat the sample, which will allow more spins to flip, which in turn further heat the sample and so a thermal runaway takes place resulting in the temperature spike and rapid change in magnetization. 

Let us consider this scenario for sample 2. To a first approximation we equate the rate that energy flows into the sample to the rate at which it can be absorbed (which increases the internal energy and raises temperature) minus the rate that the energy leaks out to the mixing chamber due to thermal conduction: 

\begin{equation*}
\dot{E}_{\rm in}= C \Delta T/dt-\dot{E}_{\rm out}
\end{equation*}

The rate of energy flowing into the sample is given by  $\dot{E}_{\rm in}= d\vec{m}/dt \cdot H$,  where $\vec{m}$ is the magnetic moment of the sample (here $\vec{m}$=M(emu/g) $\times$ mass).  Whereas the rate of energy absorbed is described by the  specific heat C $\Delta T$/dt.  The rate at which the energy flows out of the sample will be $\dot{E}_{\rm out}= \kappa S/l  \Delta T $,
where  $\kappa$  is the thermal conductivity, $S$ the surface area in thermal contact with the sample holder i.e. both 1.9~mm by 3.8~mm surfaces, $l$ is half the sample's thickness (0.45~mm), and $\Delta T$ is the temperature difference between the hot sample and the cold Cu sample holder.

\begin{figure}
\includegraphics[keepaspectratio=true, width=7.5cm]{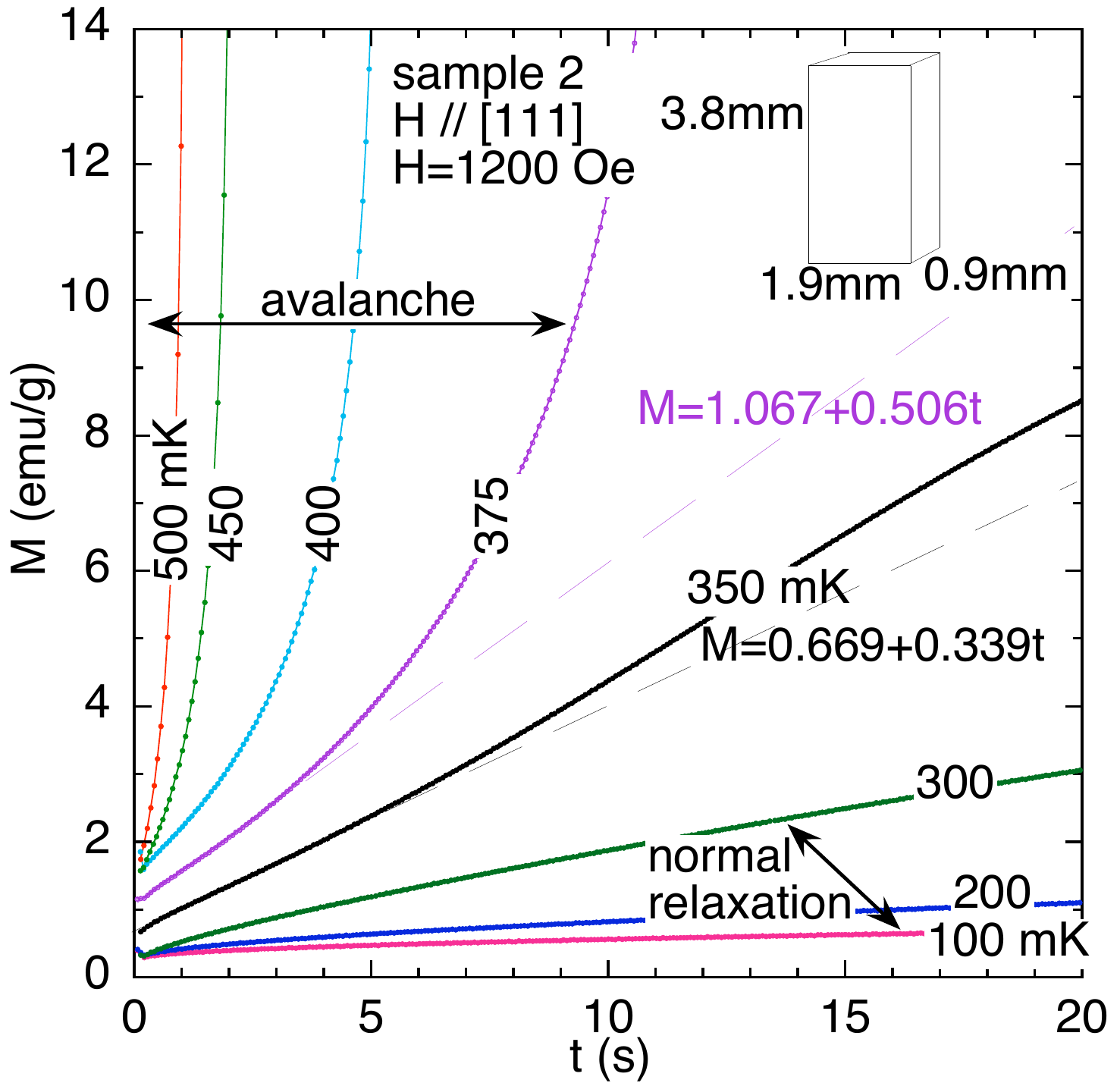}
\caption{(color online) $dM/dt$ vs time $t$ data  shown in figure \ref{dmdt_1200g} plotted on a linear time scale. The upward curvature of the magnetization as a function of time during an avalanche is clearly seen. The lines are fits to the 375 and 350 mK data }
\label{dmdt_vs_t_2}  
\end{figure}

From the 1200 Oe data, shown in Figure \ref{dmdt_1200g} and Figure \ref{dmdt_vs_t_2} on a linear time scale, it appears that the 350 mK curve is just at the limit of an avalanche, from which we obtain a value of $dM/dt=0.34$ (emu/g) $\cdot$ s$^{-1}$. Anything faster than this will trigger an avalanche. At first glance it seems reasonable to correct for demagnetization effects by using the internal field  $H_{\rm i}=H_{\rm a}-NM$. But M is small at the onset of the avalanche and we can take $H_{\rm i}\approx H_{\rm a}$. This results in  $\dot{E}_{\rm in}= 1.8$ $\mu$W. This is an extremely small power, even at 350~mK.

It is instructive to compare this power to a  rough estimate of the rate at which energy leaks out of the sample. We set the criterion that the sample temperature does not exceed 550 mK since the relaxation will be too fast at  higher temperatures, and use published values of $\kappa$ \cite{Klemke11, Kolland12} which we assume to be very similar to the thermal conductivity of our system. Thus we find
$\dot{E}_{\rm out}= \kappa S/l  \Delta T $ = 0.5 mW. This suggests that energy can leak out of the sample 270 times faster than it is getting in, thus there should be no avalanche. It therefore seems likely that energy transfer and temperature rise in the avalanche are initially confined to the spin system alone, via direct spin-spin coupling. This is consistent with the fact that the tendency to avalanche shown in Figure \ref{rlx_1200g} is suppressed below 400 mK, where the monopole gas is highly rarified and monopoles are too far apart to be influenced by each others' long ranged magnetic fields.

Note, that based on the totality of other experiments we can rule out a significant interfacial thermal resistance between the sample and the Cu holder that could reduce the magnitude of the power out of the sample to match the power in. 
Thus we conclude that the applied field, or the average internal field calculated from demagnetization corrections, can not be solely responsible for the initiation of the avalanche process. On the other hand, local fields in the pyrochlore lattice can be orders of magnitude larger than the average thermodynamic fields. In the case above, it could be possible that a small population of spins reversing in a local field of approximately 1 Tesla could be sufficient to instigate the avalanche. Another possibility is the presence of extrinsic magnetic defects which act like magnetic monopoles in the low temperature regime\cite{ccp14,Revell13}, if these magnetic excitations are participating in the thermal conduction mechanism this could, in-part explain the origin of the observed avalanches in \DyTi. It has already been demonstrated that controlling the magnetic monopole density at low temperature affects the magnetic relaxation and the subsequent approach to the equilibrium value of magnetization in both small magnetic fields which do not initiate avalanches\cite{ccp14} and in larger fields as shown in Figure~\ref{M_start=0}. The origin of the avalanches should be the basis of future work. These effects should be especially amenable to local probe measurements such as hall bar arrays and thermometers attached directly to the samples, which have been very effective in the study of avalanches in molecular nano-magnetic systems \cite{Suzuki05, McHugh07}. 

In summary, we have observed magnetic avalanches in the spin-ice compound \DyTi\ below 500 mK. This out-of-equilibrium process was observed in three different sample crystals subjected to magnetic fields along various crystallographic directions.  The fact that reversal of magnetization associated with an avalanche was observed along all measured directions suggests that avalanche is not dependent on the final spin configuration. The field at which the sample avalanches is reproducible for a given sample and field direction. However, despite being reproducible under fixed experimental conditions, the avalanche field was found to be strongly dependent on the sample and its orientation with respect to the applied magnetic field, the initial magnetization state and its history, suggesting that the avalanche field is not an intrinsic property. Its dependence on the sample geometry warrants further investigation.  Magnetic avalanches in \DyTi\ are particularly noteworthy because the process is rather slow compared to those observed in other systems. Because it is a such slow process, we were actually able to measure the magnetization during the avalanche which we describe as a self-sustained spin reversal front whose propagation is hindered by magnetic frustration.

\acknowledgments
S.R.G. greatfully acknowledges the support of the European Community — Research Infrastructures under the FP7 Capacities Specific Programme, MICROKELVIN project number 228464. M.J.J. was supported by the French ANR project CHIRnMAG.


\begin{thebibliography}{}
\bibitem{Lacroix} {\it Introduction to Frustrated Magnetism}, edited by C. Lacroix, P. Mendels, and F. Mila (Springer-Verlag, Berlin, 2011). 
\bibitem{Gardner10} J. S. Gardner, M. J. P. Gingras and J. E. Greedan, Rev. Mod. Phys. {\bf 82}, 53 (2010).

\bibitem{Siddharthan99} R. Siddharthan, B. S. Shastry, A. P. Ramirez,  A. Hayashi, R. J. Cava and S. Rosenkranz, Phys. Rev. Lett. {\bf 83}, 1854 (1999). 
\bibitem{denHertog00} B. C. den Hertog and M. J. P. Gingras {\bf 84}, 9430 (2000). 
\bibitem{Harris97} M. J. Harris, S. T. Bramwell, D. F. McMorrow, T. Zeiske, and K. W. Godfrey, Phys. Rev. Lett. {\bf 79}, 2554 (1997). 
\bibitem{Snyder04} J. Snyder, B. G. Ueland, J. S. Slusky, H. Karunadasa, R. J. Cava, and P. Schiffer, Phys. Rev. B {\bf 69}, 064414 (2004). 
\bibitem{Castelnovo08} C. Castelnovo, R. Moessner, and S. L. Sondhi, Nature {\bf 451}, 42 (2008). 
\bibitem{Ryzhkin05} I. A. Ryzhkin, I. J. Exp. and Theor. Phys. {\bf 101}, 481(2005).
\bibitem{Harris98} M. J. Harris, S. T. Bramwell, P. C. W. Holdsworth, and J. D. M. Champion, Phys. Rev. Lett. {\bf 81}, 4496 (1998). 
\bibitem{Matsuhira02} K. Matsuhira, Z. Hiroi, T. Tayama, S. Takagi, and T. Sakakibara, J. Phys. Condens. Matter {\bf 14}, L559 (2002). 
\bibitem{Sakakibara03} T. Sakakibara, T. Tayama, Z. Hiroi, K. Matsuhira, and S. Takagi, Phys. Rev. Lett. {\bf 90}, 207205 (2003). 
\bibitem{Tabata06} Y. Tabata, H. Kadowaki, K. Matsuhira, Z. Hiroi, N. Aso, E. Ressouche, and B. F\r{a}k, Phys. Rev. Lett. {\bf 97}, 257205 (2006). 
\bibitem{Fennell07} T. Fennell, S. T. Bramwell, D. F. McMorrow, P. Manuel, and A. R. Wildes, Nature Phys. {\bf 3}, 566 (2007). 
\bibitem{Moessner03} R. Moessner and S. L. Sondhi, Phys. Rev. B {\bf 68}, 064411 (2003). \bibitem{Jaubert08} L. D. C. Jaubert, J. T. Chalker, P. C. W. Holdsworth, and R. Moessner, Phys. Rev. Lett. {\bf 100}, 067207 (2008). 
\bibitem{Fennell05} T. Fennell, O. A. Petrenko, B. F\r{a}k, J. S. Gardner, S. T. Bramwell, and B. Ouladdiaf, Phys. Rev. B {\bf 72}, 224411 (2005).  
\bibitem{Ruff05} J. P. C. Ruff, R. G. Melko, and M. J. P. Gingras, Phys. Rev. Lett. {\bf 95}, 097202 (2005). 
\bibitem{Higashinaka05} R. Higashinaka and Y. Maeno, Phys. Rev. Lett. {\bf 95}, 237208 (2005). 

\bibitem{Slobinsky10} D. Slobinsky, C. Castelnovo, R. A. Borzi, A. S. Gibbs, A. P. Mackenzie, R.  Moessner, and S. A. Grigera, Phys. Rev. Lett.  {\bf 105}, 267205 (2010).  
\bibitem{Erfanifam11} S. Erfanifam, S. Zherlitsyn, J. Wosnitza, R. Moessner, O. A. Petrenko, G. Balakrishnan, and A. A. Zvyagin, Phys. Rev. B {\bf 84}, 220404(R) (2011).  

\bibitem{Macia07} F. Maci\`a, A. Hern\'{a}ndez-M\'inguez, G. Abril, J. M. Hernandez, A. Garc\'{i}a-Santiago, J. Tejada, F. Parisi, and P. V. Santos, Phys. Rev. B {\bf 76}, 174424 (2007). \bibitem{Hadjipanayis81} G. Hadjipanayis, D. J. Sellmyer, and B. Brandt , Phys. Rev. B {\bf 23},  3349 (1981). 
\bibitem{Prejean80} J.-J. Pr\'ejean, M. J. Joliclerc and P. Monod, J. Phys. Paris {\bf 41}, 427 (1980).
\bibitem{Lhotel04} E. Lhotel, C. Paulsen and A.D. Huxley, J. Mag. Mag. Mater. {\bf 272276},179 (2004). 
\bibitem{Marcano07} N. Marcano, J. C. G\'{o}mez Sal, J. I. Espeso, L. Fern\'{a}ndez Barqu\'{i}n, and C. Paulsen , Phys. Rev. B {\bf 76}, 224419 (2007). 
\bibitem{Paulsen95} C. Paulsen and J.-G.- Park, in Quantum Tunneling of Magnetization-QTM'94, edited by L. Gunther and B. Barbara, Kluwer Academic, New York, 1995, p. 171; C. Paulsen, J.-G. Park, B. Barbara, R. Sessoli and A. Caneschi, J. Mag. Mag. Mater. {\bf 140}, 1891 (1995).
\bibitem{Lhotel08} E. Lhotel, D. B. Amabilino, C. Sporer, D. Luneau, J. Veciana, and C. Paulsen, Phys. Rev. B {\bf 77}, 064416 (2008). 
\bibitem{Uehara86} M. Uehara, B. Barbara, B. Dieny, and P. C. E. Stamp, Phys. Lett. A {\bf 114}, 23 (1986).
\bibitem{Bak88} P. Bak, C. Tang, and K. Wiesenfeld, Phys. Rev A {\bf 38}, 364 (1988). 
\bibitem{Dahmen96} K. Dahmen and J. P. Sethna, Phys. Rev. B {\bf 53}, 14872 (1996). 
\bibitem{Lhotel} E. Lhotel, PhD Thesis, Universit\'e Joseph Fourier-Grenoble, France (2004). 
\bibitem{Hernandez05} A. Hern\'andez-M\'inguez, J. M. Hernandez, F. Maci\`a, A. Garc\'ia-Santiago, J. Tejada, and P.V. Santos, Phys. Rev. Lett. {\bf 95}, 217205 (2005). 
\bibitem{Suzuki05} Y. Suzuki, M. P. Sarachik, E. M. Chudnovsky, S. McHugh, R. Gonzalez-Rubio, N. Avraham, Y. Myasoedov, E. Zeldov, H. Shtrikman, N. E. Chakov, and G. Christou, Phys. Rev. Lett. {\bf 95}, 147201 (2005). 

\bibitem{Prabhakaran11} D. Prabhakaran and A.T. Boothroyd, J. Cryst. Growth {\bf 318}, 1053 (2011). 
\bibitem{Paulsen01} C. Paulsen, in Introduction to Physical Techniques in Molecular Magnetism: Structural and Macroscopic Techniques, Yesa 1999,
edited by F. Palacio, E. Ressouche, and J. Schweizer, Servicio de Publicaciones de la Universidad de Zaragoza, Zaragoza, 2001, p. 1.
\bibitem{Aharoni98} A. Aharoni, J. Appl. Phys. {\bf 83}, 3432 (1998). 
\bibitem{Osborn45} J. A. Osborn, Phys. Rev. {\bf 67}, 351 (1945). 
\bibitem{Bovo13} L. Bovo, L. D. C. Jaubert, P. C. W. Holdsworth, and S. T. Bramwell, J. Phys. Condens. Matter {\bf 25},  386002  (2013). 
\bibitem{ccp14} C. Paulsen, M. J. Jackson, E. Lhotel, B. Canals, D. Prabhakaran, K. Matsuhira, S. R. Giblin, and S. T. Bramwell,  Nature Physics {\bf 10}, 135 (2014).
\bibitem{mostame} S. Mostame, C. Castelnovo, R. Moessner, and S. L. Sondhi, PNAS {\bf 111}, 640 (2014).
\bibitem{Matsuhira11} K. Matsuhira, C. Paulsen, E. Lhotel, C. Sekine, Z. Hiroi, and S. Takagi, J. Phys. Soc. Jap. {\bf  80}, 123711(2011).
\bibitem{Ramirez99} A. P. Ramirez, A. Hayashi, R.J. Cava, R. Siddharthan, and B. S. Shastry, Nature {\bf 399}, 333 (1999). 
\bibitem{Klemke11} B. Klemke, M. Meissner, P. Strehlow, K. Kiefer, S.A. Grigera, and D.A. Tennant, J. Low. Temp. Phys. {\bf 163}, 345 (2011).
\bibitem{Kolland12} G. Kolland, O. Breunig, M. Valldor, M. Hiertz, J. Frielingsdorf, and T. Lorenz, Phys. Rev. B {\bf 86}, 060402(R) (2012). 

\bibitem{Revell13} H. M. Revell, L. R. Yaraskavitch, J. D. Mason, K. A. Ross, H. M. L. Noad, H. A. Dabkowska, B. D. Gaulin, P. Henelius, and J. B. Kycia, Nature Physics {\bf 9}, 34 (2013).

\bibitem{McHugh07} S. McHugh, R. Jaafar, M. P. Sarachik, Y. Myasoedov, A. Finkler, H. Shtrikman, E. Zeldov, R. Bagai, and G. Christou, Phys. Rev. B {\bf 76}, 172410 (2007). 

\end{thebibliography}
\end{document}